# Myasthenia Gravis Diagnosis with Surface-enhanced Raman Spectroscopy

Iuliia Riabenko[1]*, Serhiy Prokhorenko[2], Elena Klimova[3], Konstantin Beloshenko[1].

[1]*School of radiophysics, biomedical electronics and computer systems, V. N. Karazin Kharkiv National University, Freedom Square, 4, Kharkiv, 61000, Ukraine.*

[2]*Faculty of Mathematics and Applied Physics, Rzeszow University of Technology, al. Powstancow Warszawy 8., Rzeszow, 35-959, Poland.*

[3]*Laboratory of molecular and cellular technologies, Zaycev V.T. Institute of General and Emergency surgery of the National academy of medical sciences of Ukraine, Balakireva, 1, Kharkiv, 61018, Ukraine.*

*jriabenko@karazin.ua

**Abstract.** This study investigates the application of surface-enhanced Raman spectroscopy (SERS) as a diagnostic tool for myasthenia gravis, a severe neuromuscular disorder characterized by muscle weakness and fatigue. Blood serum samples were analyzed using SERS substrates embedded with 5 nm gold nanoparticles, resulting in significant Raman signal enhancement and the detection of disease-specific spectral features. The study demonstrates the potential of SERS to identify subtle biochemical changes associated with oxidative stress and altered molecular composition in serum, providing a sensitive and non-invasive approach for diagnosis. The findings highlight the method's ability to address challenges associated with traditional diagnostic tools and its potential to pave the way for more precise and early-stage detection of myasthenia gravis. This research establishes SERS as a promising complementary technique in molecular diagnostics.

Keywords: SERS, myasthenia gravis, oxidative stress, serum diagnostics, biosensors.

## Introduction

Myasthenia gravis represents the most prevalent neuromuscular junction disorder, posing challenges in diagnosis and contributing to elevated mortality rates [1], yet its understanding remains incomplete [2, 3]. Diagnostic complexities arise primarily during the disease's initial phases, characterized by isolated symptoms or localized weakness [4], progressing proximally to distal muscle weakness [5], with varying specific symptoms. Disease progression may lead to respiratory failure in over 11% of patients, resulting in painful deaths [1, 6], while other deaths are often linked to comorbidities potentially triggered by myasthenia gravis [6]. Myasthenia gravis is classified as an autoimmune disorder, characterized by the production of autoantibodies targeting proteins at the neuromuscular junction [7]. However etiological understanding of the disease remains limited [1], and given the possibility that the prevalence of myasthenia gravis is higher than currently estimated, ongoing efforts are focused on developing diagnostic guidelines that are grounded in observable clinical signs [8].

The current standard for diagnosing myasthenia gravis encompasses a combination of clinical assessments, including visual evaluations and diagnostic scales, alongside serological, electrophysiological, and pharmacological methods. However, these diagnostic approaches exhibit varying degrees of inaccuracy. Testing for acetylcholine receptor antibodies is a widely employed serological method when clinicians observe diplopia, facial ptosis, or muscle weakness. Its advantages lie in its minimally invasive nature and its ability to confirm ocular or generalized myasthenia gravis. Nevertheless, the clinical sensitivity of this test ranges from 50% to 70% [9], limiting its applicability, particularly in the early stages of the disease or seronegative cases. Single-fiber electromyography is considered the "gold standard" for confirming the diagnosis of myasthenia gravis and is highly accurate in cases of diplopia [10]. However, only 50% of patients with myasthenia gravis present with ptosis and isolated ocular symptoms as initial manifestations [11], complicating early diagnosis, as is the case with the advanced automated OMGMed system for diagnosing purely ocular myasthenia gravis [57 Li et al. (2024)]. Pharmacological tests, such as the proserin test, are valuable for confirming the diagnosis in later stages or in cases of generalized myasthenia gravis [12]. These tests provide rapid results; however, they are unsuitable for early disease detection. Although their low specificity might be addressed in the future through machine learning (ML) approaches [58 Xu et al. (2024)]. Early-stage diagnosis of myasthenia gravis remains challenging due to nonspecific symptoms and the limitations of existing methods, highlighting the need for the development of additional diagnostic tools.

In the context of SERS diagnostics, understanding the electromagnetic field enhancement factor is critical for optimizing signal amplification and detecting subtle biochemical changes in blood serum samples. High-quality SERS platforms are documented to achieve enhancement factors of at least $10^6$, with some advanced systems reaching up to $10^8$ under optimal conditions [27. Le Ru et al., 2007]. These factors depend on the surface properties of the substrate, the size and shape of nanoparticles, and the proximity of analyte molecules. A critical aspect is the reduction in nanoparticle size and the enhancement of vibrational interactions between the substrate and the sample, which position SERS as a nanoscale and single-molecule analytical biochemical tool [17, 18]. While most previous diagnostic studies using SERS have focused on oncology and cellular detection, our study broadens its application to neurology, specifically exploring its potential for diagnosing myasthenia gravis. Recent studies have also investigated the potential of infrared (IR) spectroscopy enhanced ML algorithms [59/ Severcan, F.,] for the rapid and accurate diagnosis of myasthenia gravis. This approach demonstrates high sensitivity and accuracy but requires complex computational infrastructure and a large dataset for training ML models.

By providing new insights into the poorly understood processes of oxidative stress, which lead to subtle and often difficult-to-detect changes in the molecular composition of blood serum, our study establishes a foundation for developing more precise and sensitive diagnostic methods. Utilizing substrates with stable optical properties, SERS holds significant promise as a complementary diagnostic tool that addresses the limitations of current methodologies while paving the way for future advancements.

## Material and methods

To ensure stability and accuracy in SERS data interpretation, we carefully controlled all parameters influencing signal amplification. Key factors included the filling factor of the medium [21, 22], nanoparticle concentration, and the resonance condition at the Froehlich frequency, which depends on the shape and radius of the nanoparticles [19, 20]. Maintaining fixed nanoparticle coordinates was particularly important to ensure stable field localization. In this study, we utilized substrates with implanted gold nanoparticles embedded within the fused silica surface layer, rather than synthesized nanoparticles in solution. The optical properties of these substrates were extensively characterized, confirming their ability to deliver consistent signal amplification across multiple uses with various biological samples [23, 24].

### *Manufacturing resonant structures*

Although various methods exist for fabricating SERS substrates [27, 28], the substrates used in this study were prepared following a previously described technique [23]. In brief, for the fabrication the biosensor substrate, silicon dioxide surfaces were cleaned with a potassium dichromate solution and treated with ion discharge in a vacuum to enhance adhesion [26]. Gold films were deposited onto the substrates and then subjected to laser thermal treatment, creating a temperature gradient to form nanoparticles. In the central zone, gold implantation into the surface layer of the substrate was observed. This process results in stable and reproducible substrates with well-characterized optical properties, making them suitable for SERS applications. A comprehensive description of the manufacturing process and characterization data can be referred to in [23]. The filling factor of the substrate sublayer with gold optical films was investigated using AFM, revealing a factor of $q = 0.01$. Quartz substrates were chosen because the Raman shift of quartz crystal lattice falls within the range of 300 – 500 cm$^{-1}$ [29]. The Raman scattering band within the wave number range of 600 to 700 cm$^{-1}$ is attributed to the incident wave scattering through gold nanoparticles, where plasmon oscillations are excited at a frequency of $\omega_p = 9.02$ eV [30]. This mechanism facilitates the detection of Raman shifts exceeding 500 cm$^{-1}$.

***Preparation of biological specimens***

The study involved 120 participants, including 16 female and 14 male patients with myasthenia gravis (average age: females 50±4.2 years, males 47±4.2 years), 17 female and 13 male patients with thymoma-associated myasthenia gravis (average age: females 48±4.4 years, males 45±4.1 years), and 30 female and 30 male participants in the control group (average age: females 49±5.7 years, males 46±5.6 years). Blood sampling was carried out from patients undergoing outpatient treatment at the Zaycev V. T. Institute of General and Emergency Surgery of the National Academy of Medical Sciences of Ukraine in accordance with the Helsinki Declaration of the World Medical Association. The studies were approved by the Ethics Committee of Zaycev V. T. Institute of General and Emergency Surgery.

In order to produce serum, fasting blood was collected from the ulnar vein and placed in BD-Vacutainer tubes containing Na-heparin. Each sample consisted of two tubes with a volume of 6–8 mL each. Serum was separated immediately after blood collection through centrifugation using an SM-5 MIKROMED centrifuge (2000 g, 15 minutes). The liquid component (serum) was transferred into clean polypropylene tubes using a Pasteur pipette and subsequently divided into three aliquots of 1.5 mL each.

The aliquots were frozen at –20°C in sterile 5.0 mL amber Eppendorf tubes to prevent light exposure and ensure sample stability. Before spectral analysis, the aliquots were thawed and analyzed at intervals of 5, 10, and 20 days after freezing. For each interval, one aliquot was used to prepare a single drop of serum, which was placed on the substrate and analyzed using Raman spectroscopy. This procedure ensured that three drops per participant were measured, corresponding to different storage intervals. For the control group, the samples underwent the same process, including freezing, storage, and analysis at the same intervals (5, 10, and 20 days) to maintain consistency across all experimental groups.

***Raman Spectroscopy***

Sample preparation involved depositing small volumes of serum (2 μL) onto the substrate after thawing at 20°C, followed by drying for 12 minutes using a VUP-5M vacuum evaporator at a residual pressure of 1.33 Pa, based on the principle of sessile drop formation [25]. For each participant, three drops were formed, and 4 to 6 spectra were collected from each drop. Due to the coffee-ring effect, enhanced by the Marangoni effect [link] under the specified drying conditions, proteins concentrated at the edges of the drops, defining the measurement region. Raman spectroscopy measurements were performed at various locations along the drop edges, over the gold implantation zone. Spectra collected from the central regions of the drops showed low-intensity bands and were excluded from further analysis. A schematic map of the drop segmentation used for Raman measurements is shown in Fig. 2. Over the course of the experiment, 1,320 Raman spectra were recorded from blood serum samples on substrates with implanted gold nanoparticles, including 780 spectra from patients with myasthenia gravis, of which 400 were from those with thymoma-associated myasthenia gravis. Additionally, 540 spectra were recorded from the edges of drops in the control group.

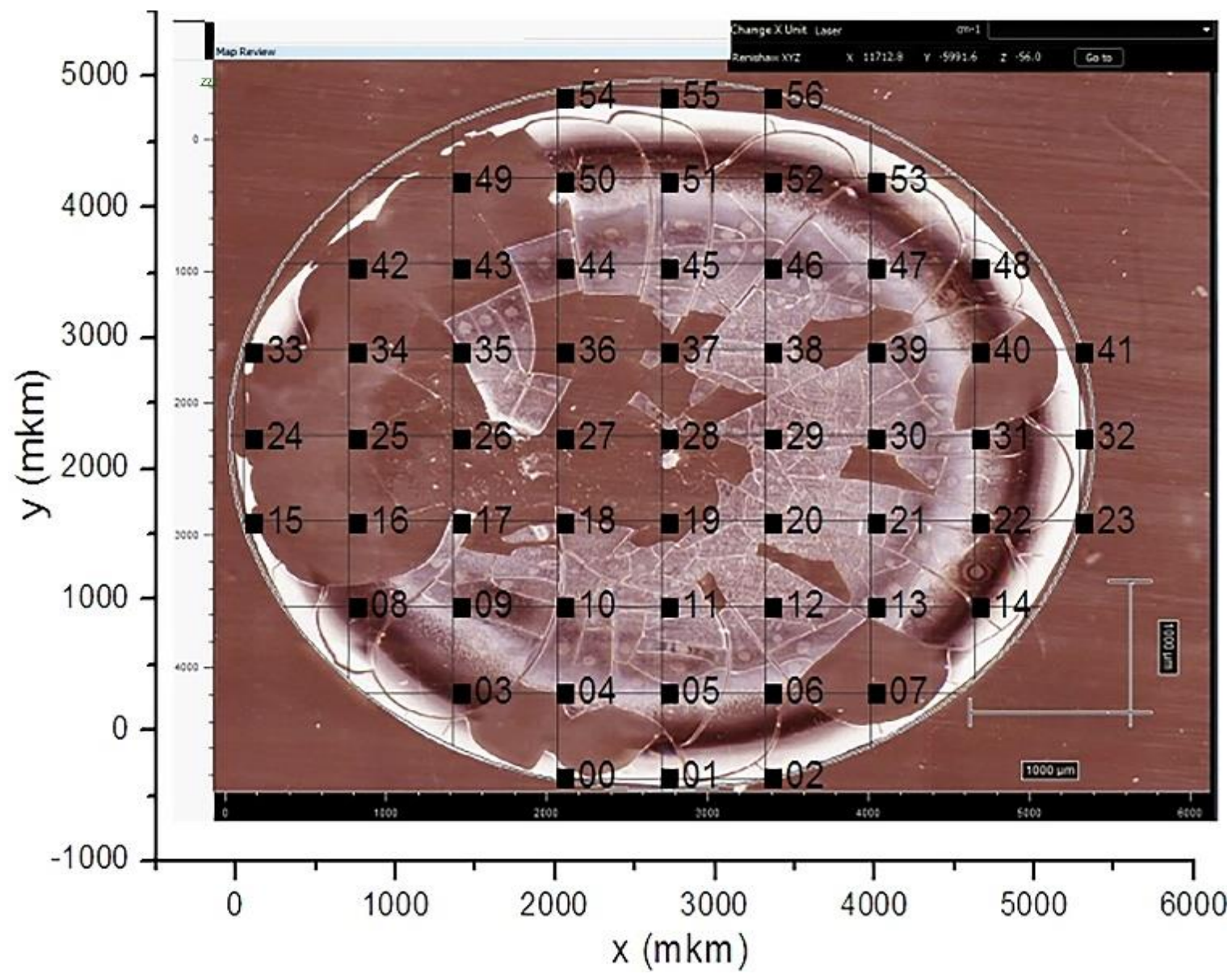

Figure 1. Segmenting blood serum drop samples to acquire spectra from the edges of the drops.

The measurement range for Raman shift analysis spanned from 150 to 3200 $cm^{-1}$, with a spectral resolution of 2 $cm^{-1}$ using the inVia confocal Raman microscope (Renishaw plc, Gloucestershire, UK). The detector employed was a Renishaw CCD camera, boasting dimensions of 1040×256 pixels. The laser source utilized was the I0785SR0090B-IS1 (Innovative Photonic Solutions, New Jersey, USA) operating at a wavelength of 785 nm (mode: normal), with a spectral width of <100 MHz (<0.2 pm, <0.005 $cm^{-1}$) and an output power of 90 mW. The laser power was set at 1%, and a grid of 1200 lines per mm (633/780) was employed. The exposure time was set at 10 s, utilizing a lens with a magnification factor of 5x. The software utilized for spectrum acquisition was WiRE™ 3.4, followed by processing using WiRE™ 5.6 (Renishaw plc, Gloucestershire, UK).

## Results

Blood serum drops from patients with myasthenia gravis, with or without thymoma, are depicted in Fig. 1, captured using the Zeiss™ Primo Star HAL LED microscope (Carl Zeiss Microscopy GmbH, Jena, Germany). The mechanism of organic molecules distribution in drops can be described based on their molar mass. The Gorsky effect [31] of upward diffusion is observed when the drop is drying. Molecules with a higher molar mass are then accumulated at the periphery of the drop. Water and other light components evaporate while drying which shows where to focus the laser beam and how the microscope system should be set up to successfully detect the spectra. Additionally, microphotographs of dried drops of blood serum have been added to the supplemental material.

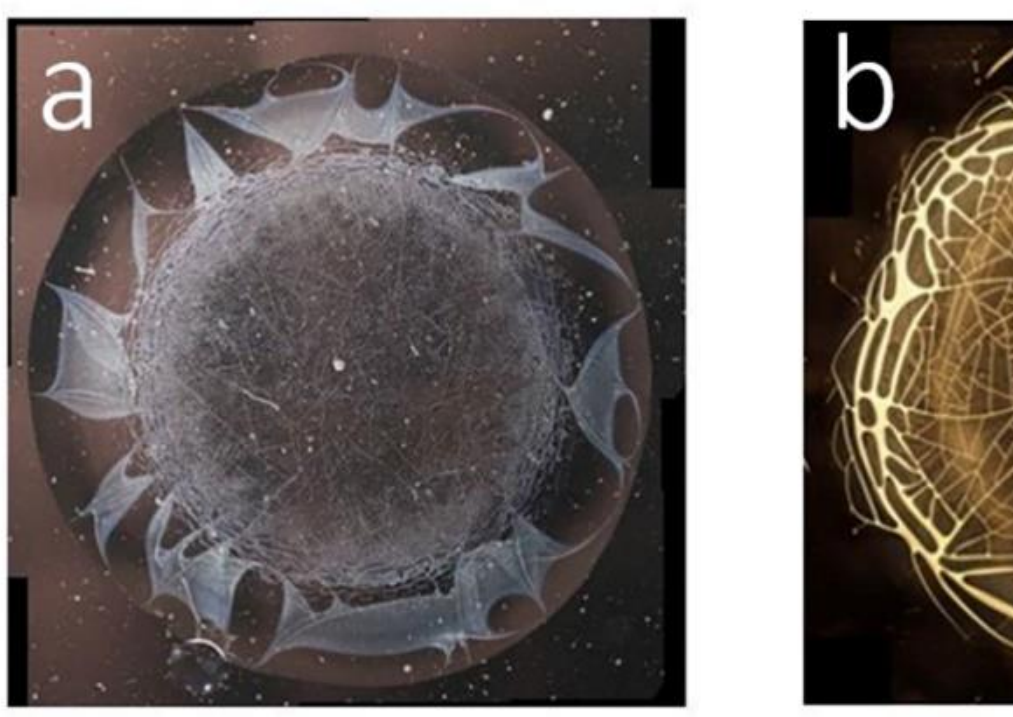

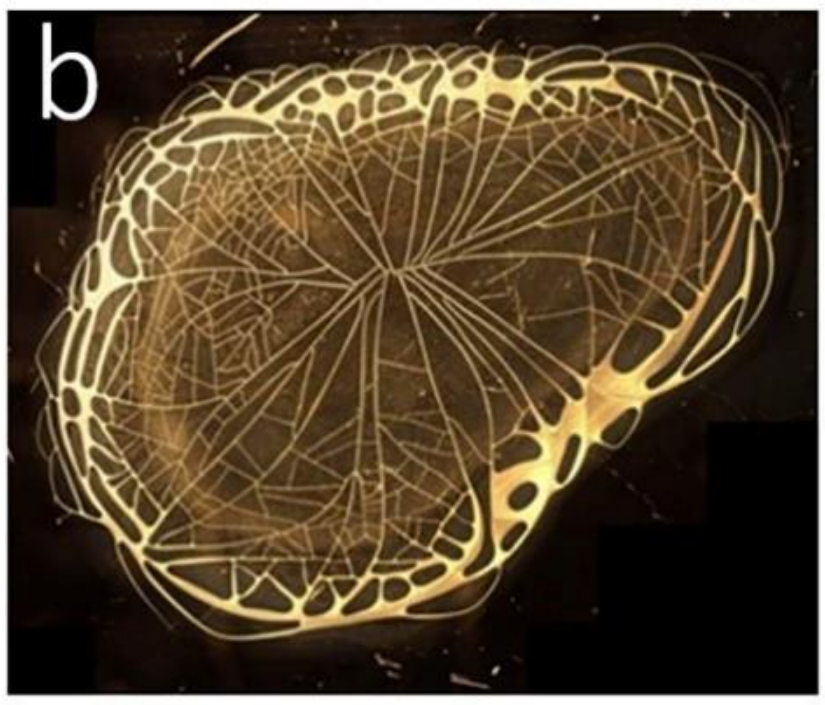

Figure 1. Micrographs of blood serum drops: a) patient with thymoma-associated myasthenia gravis; b) patient with myasthenia gravis.

The gold nanoparticles implanted into $SiO_2$, situated at a depth of 5 nm beneath the flat surface of the fused quartz glass, constitute a high-quality resonance system ($Q \sim 0.8 - 0.9$) with a resonance plasmon frequency of $3.31 \times 10^{15}$ $s^{-1}$ [25]. These nanoparticles create a system of coupled oscillators with the serum monolayer [32], resulting in enhanced Raman scattering [33] from polarizable groups in organic molecules.

Raman data have been processed by software WiRE 5.6 that allowed to remove of noise interference and oversaturated spectra. The non-informative part of the spectrum was truncated. Cosmic ray removal was performed across the entire spectrum, with the bandwidth set to 3 – 7 pixels. The autofluorescent background was eliminated using an intelligent spline node set to 11. Then the Savitzky–Golay filter was used for spectra smoothing by a 3rd degree polynomial function. The normalization of the spectrum before further analysis and a comparison of different spectra has been applied. The normalization and subtraction of the substrate spectrum was conducted using identical experimental parameters (exposure time and sample position) as the main measurements.

The supplemental material contains an example of a Raman spectrum obtained from the edge of a blood serum drop taken from patients diagnosed with myasthenia gravis. The spectra shifted for ease of comparison. After completing all processing stages, the spectrum was analyzed based on its position and peak intensity. The WiRE 5.6 software identified peaks in the active spectrum using the basic threshold settings.

The resulting spectrum of the substrate is presented in the supplemental material and result aligns well with existing data [34]. The average spectra of blood serum (Fig. 3) were obtained by calculating and analyzing processed data using OriginPro 8.0 software (OriginLab, Massachusetts, USA), similar to the method described in [35]. To assess the sensitivity and specificity of the method, a comparative analysis of spectra from patients with myasthenia and a control group was performed, which allowed for the identification of disease-specific characteristic peaks.

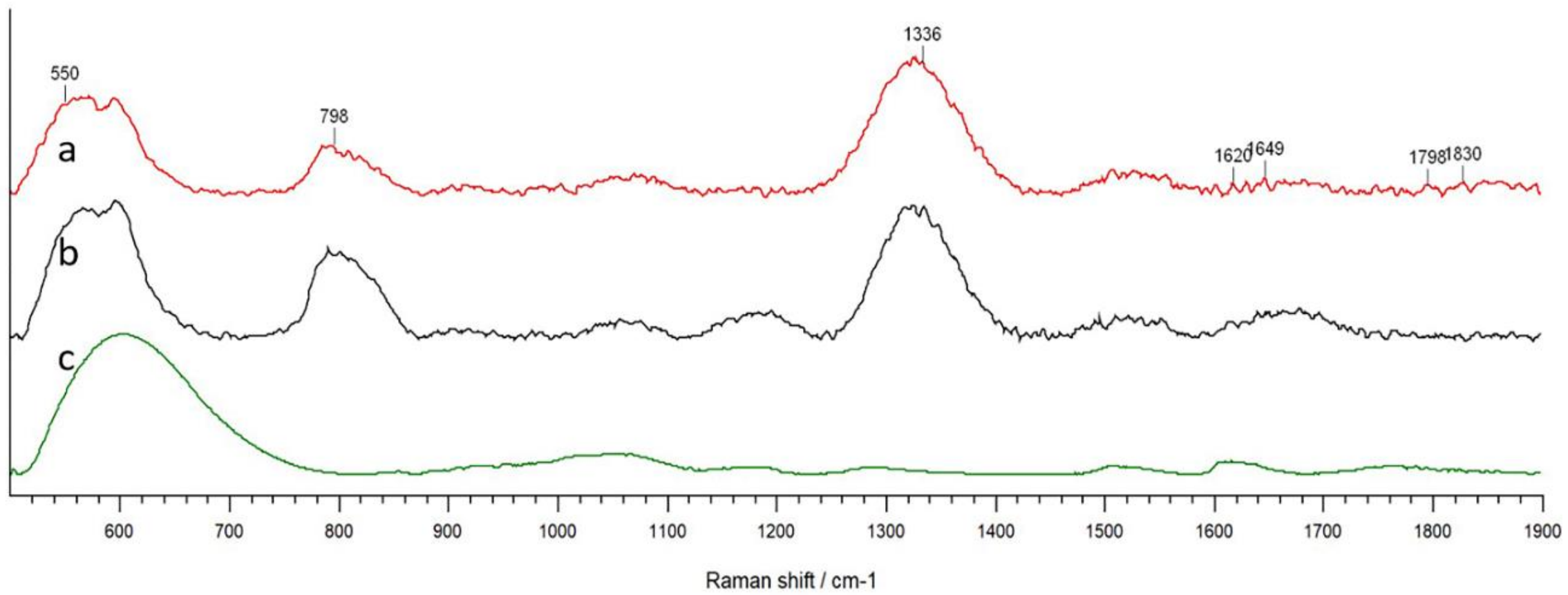

Figure 3. Average Raman spectra in blood serum were observed in: a) patients with myasthenia gravis; b) those with thymoma-associated myasthenia gravis; and c) healthy research participants.

The data presented in the average spectrum (Fig. 3) reveal a high intensity of the fundamental modes of the NO2 molecule. To ensure clarity in the data interpretation, the spectrum of the sample before processing, the substrate spectrum, and the processed spectrum after subtraction are shown in the supplementary materials. Importantly, no significant differences were observed between the spectra of patients with thymoma-associated myasthenia gravis and those with myasthenia gravis without thymoma association. This suggests that similar oxidative stress processes occur in both groups. While minimal differences were noted, they were not the focus of this study and require further investigation in subsequent research. To quantify the occurrence of other peaks in spectroscopic data collected from all samples, it involved calculating the frequency of specific peak occurrences (Fig. 4) across the samples, using Python Programming Language (PSF, Delaware, USA).

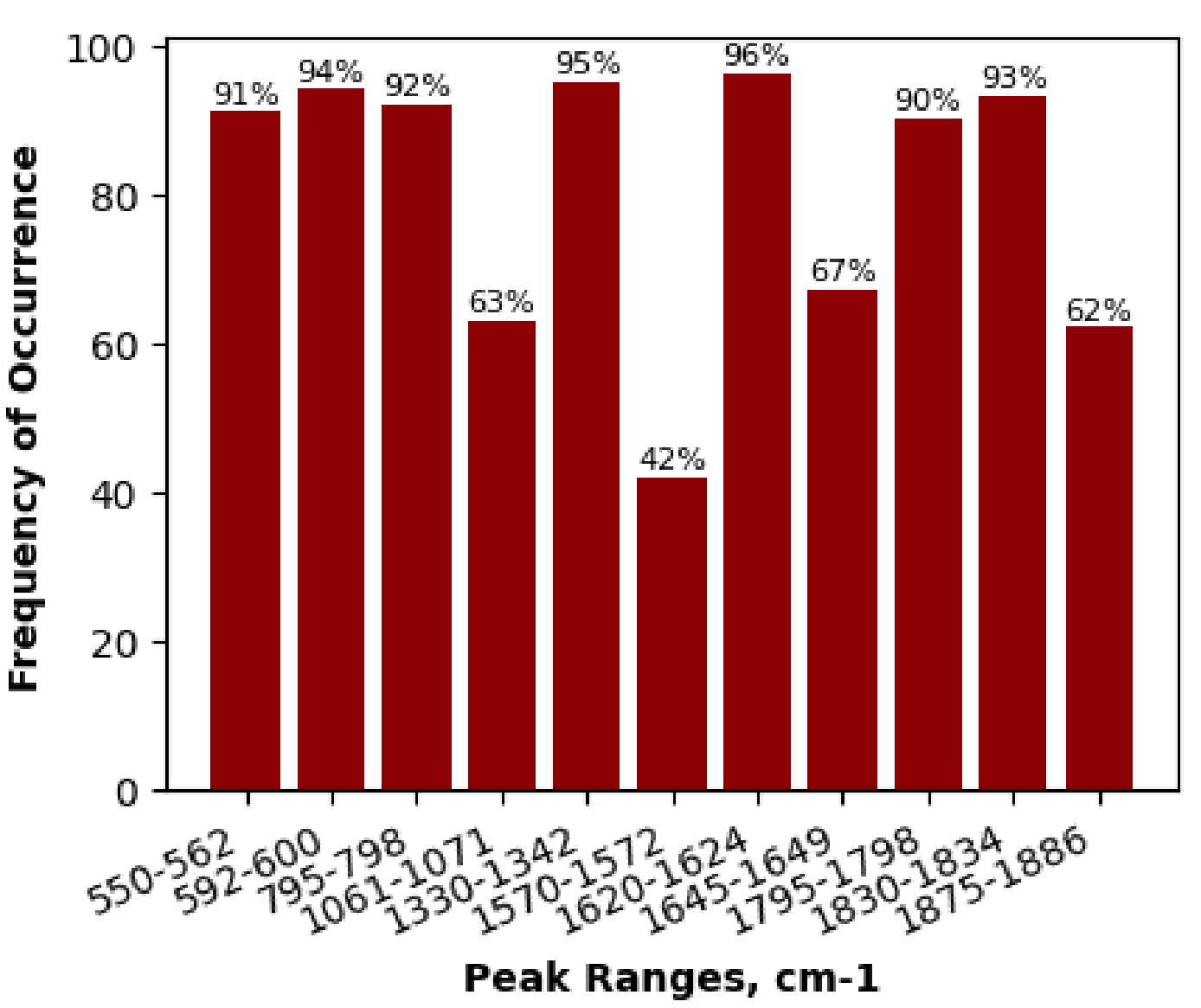

Figure 4. Frequency analysis of specific peak occurrences.

The statistical frequency analysis indicates that these measurements were consistently stable. This observation is evident both in terms of peak presence in the samples and the narrow range of peak detection, which ensured to the stability of the SERS method due to the implantation of nanoparticles.

**Discussion**

Raman scattering is enhanced due to the two-photon process of interaction between the energy of the polariton field, laser radiation, and vibrational symmetric modes. The enhancement of Raman scattering can be estimated as follows [36]:

$$P_{SERS} = G_{SERS}P_{Raman} = G_{SERS}^{Em}G_{SERS}^{Chem}R_{Raman} =$$

$$= G_{SERS}^{Em}G_{SERS}^{Chem}N\alpha_R(\omega_R, \omega_L)E(\omega_L), \quad (1)$$

where $G_{SERS}$ - coefficient that takes into account the amplification caused by the substrate, which consists of two electromagnetic coefficient $G_{SERS}^{Em}$ and chemical $G_{SERS}^{Chem}$

Describing the gain factors we get:

$$P_{SERS} = \frac{\omega_R^4}{32\pi\varepsilon_0 c^3}\frac{|E_{LOC}(\omega_L)|^2}{|E(\omega_L)|^2}\frac{|E_{LOC}(\omega_R)|^2}{|E(\omega_R)|^2}N|\alpha_R(\omega_R, \omega_L)E(\omega_L)|^2\frac{\sigma_k^{das}}{\sigma_k^{free}}. \quad (2)$$

The effective cross section of Raman scattering of a molecule deposited on a substrate can be described by the equation [37]:

$$\sigma_\alpha(r, \omega) = \frac{16\pi^2\omega_{fi}}{cE_0^2}|\langle f|H(r, \omega)|i\rangle|^2 \times \delta\left[h(\omega - \omega_{fi})\right] \sim \sigma_\alpha^{(0)}\left|3\frac{\varepsilon'(\omega)}{\varepsilon''(\omega)}\right|^2|h(r)|^2, \quad (3)$$

where $\sigma_\alpha^{(0)}$- is the cross section of free molecules in the absence of metal, $i \rightarrow f$ is the optical transition from state $i$ to $f$, $\langle f|H(r, \omega)|i\rangle$ - is the matrix Hamiltonian of the interaction of a molecule with a field, $h(r)$ is the gain of the dipole moment of the adsorbed molecule. Then

$$P_{SERS} = \frac{\omega_R^4}{12\pi\varepsilon_0 c^3}\frac{|E_{LOC}(\omega_L)|^2}{|E(\omega_L)|^2}\frac{|E_{LOC}(\omega_R)|^2}{|E(\omega_R)|^2}N|\alpha_R(\omega_R, \omega_L)E(\omega_L)|^2\left|3\frac{\varepsilon'(\omega)}{\varepsilon''(\omega)}\right|^2|h(r)|^2. \quad (4)$$

In the case of Au [25] $|\varepsilon'(\omega)|/\varepsilon''(\omega) \approx 10^3$, which is an order of magnitude greater than that of other noble metals. It is this fact that determines the choice of gold as the basis of nanoparticles for enhancing Raman scattering by biomolecules.

If the distance between the molecular group and metal granules falls within the range of 20 – 35 nm [32], we can infer the existence of an additional component of polariton energy during Raman scattering of the molecular group. This suggests the presence of a system of coupled oscillators. On the other hand, it is known that the electric field strength in granules is directed perpendicular to the curvature (Fig. 5 a, b) of the metal surface. Thus, the modes of vibration of molecular groups lying in the plane of the substrate will experience maximum amplification. Such modes are the mode of symmetric stretching of the molecular group (Fig. 5).

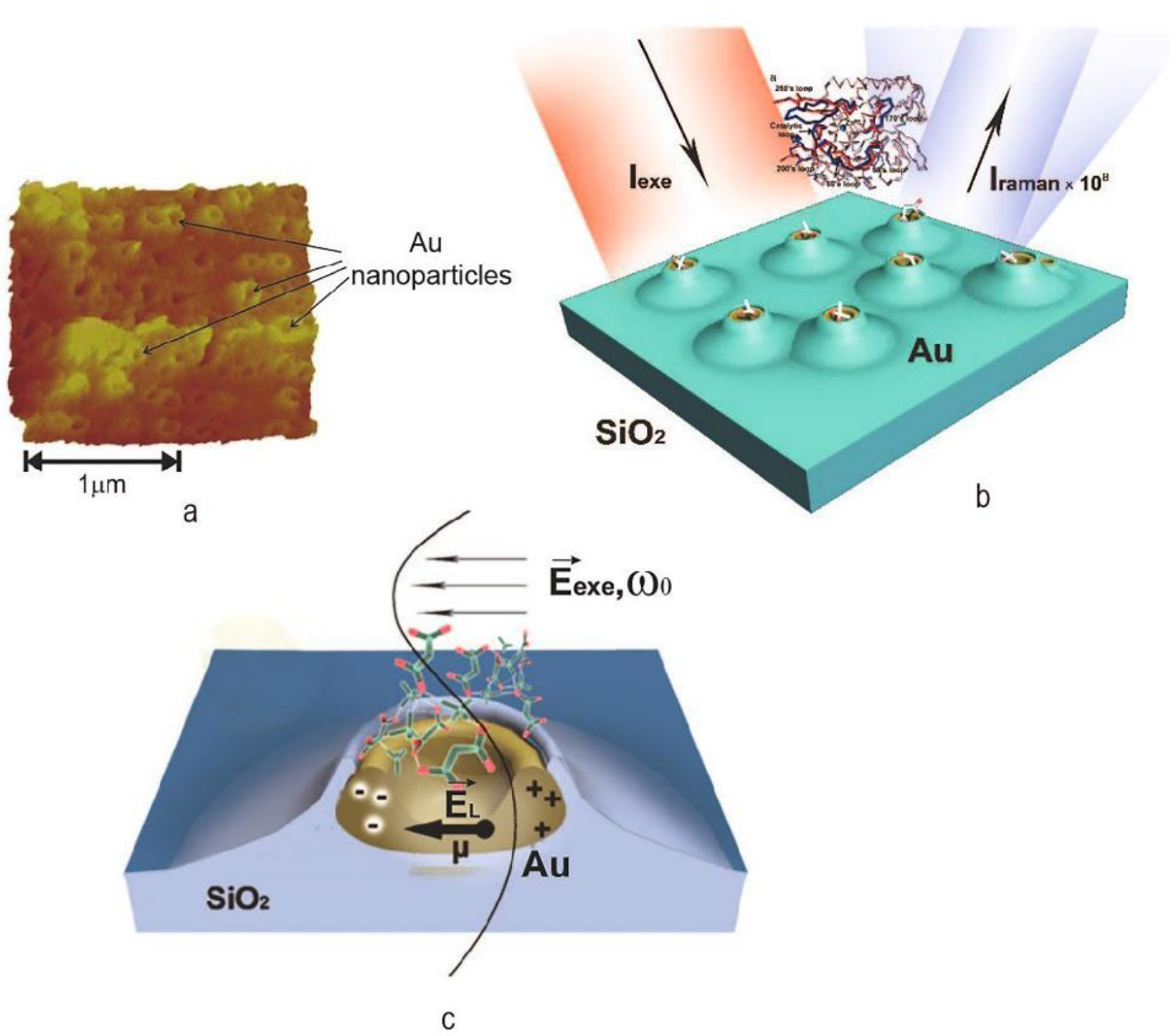

Figure 5. Enhancement of Raman scattering facilitated by implanted nanoparticles: a) AFM image of Au nanoparticles implanted in $SiO_2$; b) scheme illustrating combined amplification by metal nanoparticles and their interaction with sample components; c) molecular groups of molecules within the vicinity of metal granules.

Both the presence of a peak and its position are crucial in the study of Raman spectra, as the polarizability tensor of a molecule is contingent upon the solution or medium containing the detected component. Therefore, conducting a comparative analysis of the Raman spectra obtained using the WiRE 5.6 spectral database with those sourced from literature [38] could suggest the presence of Botulinum toxin A spectra in the samples. The obtained spectra feature the $NO_2$ group, which is part of the molecule, exhibiting a stretching mode of the aromatic ring excited with a shift $\Delta\nu = 19$ cm$^{-1}$ and $\Delta\nu = 3$ cm$^{-1}$, respectively, due to coupling. This shift is insignificant, owing to the difference in energy utilized to alter the polarizability tensor of the medium constituents. A more substantial field localization is necessitated to augment the polarizability of the aromatic ring.

The literature demonstrates that human monoclonal antibodies effectively neutralize several serotypes of botulinum neurotoxin [39, 40], thereby reproducing the antigen binding model. Meanwhile, nitrogen dioxide is participating in numerous biological reactions [41, 42], including pathological ones and it can be detected taking into account its lifetime [43, 44]. However, it can also be identified in the composition of 3-nitrotyrosine, serving as a crucial Raman reporter molecule, which is produced through the nitration of tyrosine with nitric oxide during apoptosis [45]. Furthermore, in the analyzed averaged spectrum, the peaks at 1620 cm$^{-1}$ and 1649 cm$^{-1}$ exhibit less prominence. These peaks have been previously documented in the literature for botulinum toxin detection using the SERS method with labeling, wherein polyclonal animal antibodies were employed to create the label [46]. However, in the context of measuring a signal in humans, this approach yields a lower signal-to-blank ratio.

In view of the above and considering that biochemical investigations of blood parameters [47] and cytokine profiling of immune responses in myasthenia gravis patients have revealed molecular processes akin to those observed in botulinum toxin poisoning [48 – 51]. The accumulation of botulinum toxin in various organs, including the brain [52], can trigger primary myasthenia gravis symptoms by affecting the innervation of extraocular muscles, resulting in ptosis and paralysis. While our study does not aim to establish a causal relationship, it underscores the necessity for further investigation into the potential link between exposure to low concentrations of botulinum toxin and the development of myasthenia gravis. By emphasizing the toxin's possible roles in oxidative stress and immune dysregulation, we underline the complexity of the disease's pathogenesis, which aligns with prior clinical observations [46.50.56Punga AR, Liik M; Dell'Antonia et al.]. Although the sample size in this study was relatively small, our findings highlight the importance of the analysis conducted and the need for larger-scale studies with controlled experimental designs to validate the spectral observations presented here and to delineate the potential impact of botulinum toxin on disease onset and progression.

Previous studies have suggested a potential link between botulinum toxin exposure and the pathogenesis of myasthenia gravis, highlighting overlapping molecular processes such as oxidative stress and its impact on neuromuscular function [50.Punga AR, Liik M, 2020]. Clinical observations have demonstrated that even therapeutic or cosmetic applications of botulinum toxin type A can unmask or exacerbate symptoms of latent myasthenia gravis, reinforcing its relevance in understanding the disease's etiology [56. Dell'Antonia et al., 2023].

In the context of diagnostic methodologies, SERS-based approaches have proven effective for detecting botulinum toxin molecules due to their high sensitivity and capability for molecular-level analysis [46. Kim K, et al., 2019]. This supports the use of SERS in analyzing blood serum samples to explore the possible biochemical pathways linking botulinum toxin exposure to myasthenia gravis. While the current study identifies spectral features consistent with oxidative stress and the presence of nitro groups, it does not confirm a causal relationship between botulinum toxin and the disease. Instead, the findings provide a basis for future research, aiming to explore this hypothesis in larger patient cohorts and with controlled experimental conditions.

While both IR spectroscopy with ML and SERS target the analysis of biochemical changes associated with myasthenia gravis, their approaches differ significantly. IR spectroscopy benefits from the ability to analyze broader molecular fingerprints, which, when combined with advanced ML algorithms, allows for robust classification of disease states. However, this approach requires sophisticated computational infrastructure and a large dataset for training ML models. In contrast, SERS provides exceptional sensitivity for detecting specific molecular changes, such as oxidative stress markers and nitro groups, at ultra-low concentrations. Additionally, SERS-based methods may be more accessible for resource-limited settings due to their relatively straightforward experimental setup. Both methods demonstrate complementary strengths, and further comparative studies could elucidate their respective roles in clinical diagnostics.

Our findings not only validate the potential of using SERS for the detection of myasthenia biomarkers, but also demonstrate the broader applicability of this technique in disease diagnostics. By employing principal component analysis on the peak positions, which correspond to the vibrational modes of characteristic molecular groups that remain stable during sample handling and reflect oxidative stress, it is possible to classify specific diseases. This approach presents a promising alternative to traditional cytokine profiling.

**Conclusion**

The efficacy of utilizing Raman scattering to identify markers of myasthenia gravis has been validated. By employing implanted nanoparticles into the surface layer of fused quartz, it becomes feasible to capture signals of varying polarization, thereby demonstrating that the local field emanating from metallic gold nanoparticles enhances the Raman scattering of pathological compounds present in the blood serum of myasthenia gravis patients. It has been ascertained that spectra indicative of the presence of botulinum toxin A are discernible in the blood of these patients. The intricate structure of toxins in relation to Raman spectroscopy serves as a critical attribute, as it manifests through multiple bands, consequently heightening the reliability of the outcomes. Nevertheless, it is precisely through analyzing the position of the Raman scattering band from various molecular groups, while considering the oscillation between the medium's components, that assertions about the constituent elements can be made even without the direct "fingerprint" of entire molecules. Therefore, the above analysis of peaks and the results of the search for a known spectrum are worthy of special attention in the future. Conducting additional spectroscopic studies, in which the infection processes with botulinum toxin A will be simulated, with its controlled residual concentrations on the proposed substrates, would level out the limitations of such analysis from WiRE software. Also, based on the literature data, to determine the membership of the $NO_2$ group in future studies at different stages of myasthenia, it is essential to control ascorbic and uric acids, protein thiol groups, and the reserves of alpha-tocopherol, bilirubin and ubiquinol-10 [53]. Control of a subpopulation is also possible T cells [54] and a subpopulation of lymphocytes [55] to test the hypothesis of the immune response development mechanism put forward.

The findings of this study demonstrate stable SERS signals, which are associated with the implantation of nanoparticles in the near-surface layer of fused quartz. This configuration ensures consistent distances between the nanoparticles and the detected molecules, stability of the nanoparticles' coordinates, and allows for the repeated use of a single substrate throughout the experiment. The small size and optimized shape of the nanoparticles facilitated signal enhancement through interband transitions and the excitation of hybrid modes, achieving a level of sensitivity previously unattained in similar studies. Further confirmation of the role of botulinum toxin in the pathogenesis of myasthenia gravis and the effectiveness of the SERS method as a diagnostic tool requires additional research, including larger patient cohorts and multicenter studies. Nevertheless, the current findings underscore the significant potential of the SERS method in advancing the diagnosis and understanding of myasthenia gravis, highlighting the method's ability to deliver reliable and sensitive results that could transform clinical practices.

**In Memoriam:**

The authors gratefully acknowledge the significant contributions of Professor V. M. Shulga from the College of Electronic Science and Engineering, Jilin University, China. Professor Shulga was a distinguished expert in physics specializing in electrodynamics and metamaterials, and he played a pivotal role in guiding the theoretical framework and experimental design of this research project. Sadly, Professor Shulga passed away before the completion of this manuscript. We dedicate this article to his memory and recognize his enduring impact on our field.

**Acknowledgments:**

The authors would like to thank Dr. Eng. Dariusz Żak from the Department of Functional Materials at Rzeszow University of Technology for his valuable assistance in conducting spectral measurements for this research. His expertise and support were instrumental in the successful completion of this study.

**Declaration of Interest Statement:**

The authors declare that they have no conflicts of interest that could influence the research presented in this manuscript.